\documentclass[preprint2]{emulateapj}
\usepackage{multirow}
\newcommand{\Reff}{$r_h$}

\newcommand{\kms}{\ensuremath{\rm{km\,s^{-1}}}}

\newcommand{\MLsun}{\ensuremath{M_\odot/L_\odot}}

\shorttitle{A tidally disrupting dwarf galaxy in the halo of NGC~253}
\shortauthors{Toloba et al.}

\begin{document}

\title{A tidally disrupting dwarf galaxy in the halo of NGC~253}


\author{Elisa~Toloba\altaffilmark{1,2}}\email{toloba@ucolick.org}
\author{David~J.~Sand\altaffilmark{1}}
\author{Kristine~Spekkens\altaffilmark{3}}
\author{Denija~Crnojevi\'c\altaffilmark{1}}
\author{Joshua~D.~Simon\altaffilmark{4}}
\author{Puragra~Guhathakurta\altaffilmark{2}}
\author{Jay~Strader\altaffilmark{5}}
\author{Nelson~Caldwell\altaffilmark{6}}
\author{Brian~McLeod\altaffilmark{6}}
\author{Anil~C.~Seth\altaffilmark{7}}

\affil{$^1$Texas Tech University, Physics Department, Box 41051, Lubbock, TX 79409-1051, USA}
\affil{$^2$UCO/Lick Observatory, University of California, Santa Cruz, 1156 High Street, Santa Cruz, CA 95064, USA}
\affil{$^3$Department of Physics, Royal Military College of Canada, P.O. Box 17000, Station Forces, Kingston, Ontario, K7L 7B4, Canada}
\affil{$^4$Carnegie Observatories, 813 Santa Barbara Street, Pasadena, CA 91101, USA}
\affil{$^5$Department of Physics and Astronomy, Michigan State University, East Lansing, MI 48824, USA}
\affil{$^6$Harvard-Smithsonian Center for Astrophysics, Cambridge, MA 02138, USA}
\affil{$^7$Department of Physics and Astronomy, University of Utah, Salt Lake City, UT 84112, USA}

\begin{abstract}

We report the discovery of Scl-MM-Dw2, a new dwarf galaxy at a projected separation of $\sim$50~kpc from NGC~253, as part of the PISCeS (Panoramic Imaging Survey of Centaurus and Sculptor) project.
We measure a tip
of the red giant branch distance of $3.12\pm0.30$~Mpc, suggesting that
Scl-MM-Dw2 is likely a satellite of NGC 253.
We qualitatively compare the distribution of red giant branch (RGB) stars in the color-magnitude diagram with theoretical isochrones and find that it is consistent with an old, $\sim$12~Gyr, and metal poor, $-2.3<$~[Fe/H]~$<-1.1$, stellar population. We also detect a small number of asymptotic giant branch stars consistent with a metal poor $2-3$~Gyr population in the center of the dwarf. Our non-detection of HI in a deep Green Bank Telescope spectrum implies a gas fraction $M_{HI}/L_V<0.02$~\MLsun. The stellar and gaseous properties of Scl-MM-Dw2 suggest that it is a dwarf spheroidal galaxy. Scl-MM-Dw2 has a luminosity of $M_V=-12.1\pm0.5$~mag and a half-light radius of \Reff~$=2.94\pm0.46$~kpc which makes it moderately larger than dwarf galaxies in the Local Group of the same luminosity.
However, Scl-MM-Dw2 is very elongated ($\epsilon=0.66\pm0.06$) and it has an extremely low surface brightness ($\mu_{0,V}=26.5\pm0.7$~mag~arcsec$^{-2}$). Its elongation and diffuseness make it an outlier in the ellipticity-luminosity and surface brightness-luminosity scaling relations. 
These properties suggest that this dwarf is being tidally disrupted by NGC~253.

\end{abstract}

\keywords{galaxies: individual (NGC~253) --- galaxies: dwarf --- galaxies: stellar content --- galaxies: halos --- galaxies: photometry}

\section{Introduction}

The $\Lambda$+Cold Dark Matter
($\Lambda$CDM) model for structure formation is very successful on
large scales ($\gtrsim$10 Mpc), and within this model galaxies grow hierarchically by accreting smaller dark matter halos
\citep{Springel06}.  
Models of mass assembly based on the $\Lambda$CDM paradigm predict a wide range of halo substructure -- dwarf galaxies and stellar streams with varying numbers and morphology due to different accretion histories \citep[e.g.][]{Johnston08,Cooper10}.   

A major observational effort is underway to identify, quantify, and analyze halo substructures in order to directly compare them with expectations from simulations -- even in environments beyond the Local Group.  For instance,  \citet{Chiboucas09,Chiboucas13} have searched the M81 group of galaxies, in resolved stars, and find fourteen satellites down to a magnitude of $M_V$$\approx$$-$10. Likewise, a search for dwarf companions to the dwarf galaxy NGC~3109 has yielded at least one new satellite with $M_{V}$=$-$9.7 \citep{AntliaB}.  
Low surface brightness searches have yielded many new dwarf satellites \citep[e.g.][seven found around M101]{Merritt14}, as well as streams and other halo substructure \citep{MartinezDelgado10,MartinezDelgado12,MartinezDelgado14}.  

Here we report the discovery of a faint, diffuse and highly elongated satellite galaxy located $\sim$50~kpc in projection from NGC~253, which we dub Scl-MM-Dw2, in accordance with our prior work in the Sculptor group. The data were taken as part of the Panoramic Imaging Survey of Centaurus and Sculptor \citep[PISCeS;][]{Sand14,Crnojevic14}, which aims to study the faint satellites and stellar halo substructure of NGC~253 and NGC~5128 in resolved stellar light.  Due to its high ellipticity and faint hints of other nearby substructure, Scl-MM-Dw2 is likely being disrupted, allowing for a direct view of the continuing buildup of NGC~253's halo.  In Section \ref{observations}, we describe our data and reduction methods, along with the procedure followed to discover the dwarf galaxy. In Section \ref{galaxy}, we measure the physical properties of the new dwarf, estimate its distance, and study its location in the scaling relations of other dwarf galaxies. In Section \ref{discussion}, we discuss our findings, summarize the main properties of Scl-MM-Dw2, and conclude.

For reference, $1' =1.0$~kpc at our assumed distance to NGC~253 \citep[3.47~Mpc;][]{RadburnSmith11}.

\section{Observations and Discovery of Scl-MM-Dw2}\label{observations}

\begin{figure}
\centering
\includegraphics[angle=0,width=10cm]{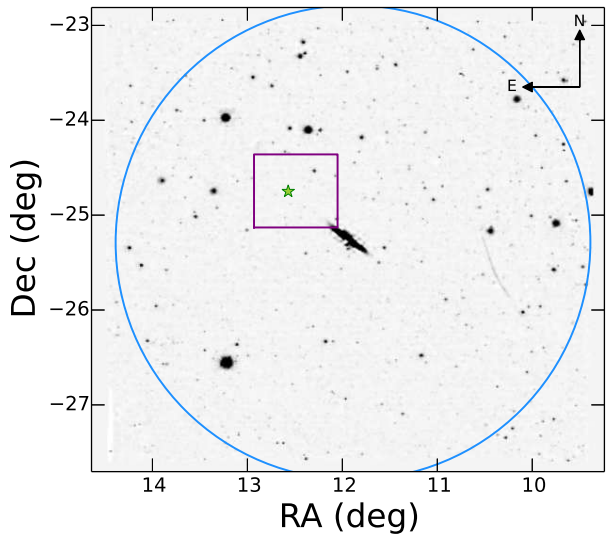}
\caption{DSS image centered on NGC~253. The blue circle indicates the planned extent of the PISCeS survey (R=150~kpc). The purple box indicates the region discussed here. The green star indicates the position of Scl-MM-Dw2.}\label{DSSmap}
\end{figure}

The data presented here are part of the larger PISCeS survey, which utilizes the Megacam instrument \citep{McLeod15} located at the $f/5$ focus of the Magellan Clay 6.5~m telescope (Las Campanas Observatory, Chile). Megacam has a $24'\times24'$ field of view and a binned pixel scale of $0\farcs16$.  Here we focus on the four Megacam fields in the vicinity of Scl-MM-Dw2. 
Figure \ref{DSSmap} shows their location in an extended Digitized Sky Survey image centered on NGC~253.

PISCeS typically observes each field for $6\times300$~s in the $g$ and $r$ band to achieve image depths of $\sim$26-27 mag, although longer exposure times are necessary in poor weather and/or seeing conditions.  The four fields discussed here were observed separately between November 2011 and October 2014, with final image point spread functions between 0\farcs65 and 1\farcs0. 
Initial data reduction, which consists of image detrending, astrometric matching, and stacking of the multiple individual exposures, is performed by the Smithsonian Astrophysical Observatory Telescope Data Center using the pipeline designed for Megacam by M. Conroy, J. Roll, and B. McLeod. We perform point spread function (PSF) fitting photometry on the final stacked images using the DAOPHOT and ALLSTAR software suite \citep{Stetson87,Stetson94}, matching the $g$ and $r$ band catalogs with  DAOMATCH/DAOMASTER \citep{Stetson93}. The deepest catalogs possible are obtained by running ALLFRAME \citep{Stetson94} to perform simultaneous photometry for objects detected in the individual $g$ and $r$ band stacks. We culled our catalogs of outliers in $\chi$$^2$ versus magnitude, magnitude error versus magnitude, and sharpness versus magnitude space to remove objects that were not point sources.

To flux calibrate the data we employ a dual approach.  First, we observe equatorial fields from the Sloan Digital Sky Survey \citep[SDSS;][]{SDSSDR12} at varying airmasses on photometric nights. We transform instrumental magnitudes measured in our Megacam images into SDSS magnitudes with our derived zeropoints, color terms and extinction coefficients derived from the SDSS imaging.  For nights that are not photometric, we take advantage of the tile overlaps between Megacam pointings (typically $\sim$2 arcmin) to bootstrap our photometry across the survey from photometric nights.  The data presented in this work were flux calibrated utilizing the second approach, as none of it was taken in completely photometric conditions.  The final, flux calibrated catalogs were corrected for Galactic extinction on a star-by-star basis \citep{Schlafly11}; all magnitudes reported in this work are corrected in this way.

Once the data are photometered, we use two techniques to search for faint dwarfs and other substructure.  First, we visually inspect all the images, searching for spatially compact overdensities of stars which have some signs of diffuse light as well.  Second, we create red giant branch (RGB) stellar maps from our point source photometric catalog by selecting stars that are broadly consistent with RGB stars at the distance of NGC~253 (see Figure \ref{map}, which was made using the selection box in Figure \ref{CMD}). The main sources of contamination are foreground stars, seen as a nearly vertical sequence at $g-r\sim0.5$, unresolved background galaxies ($g-r\sim 0.0$), and NGC~253's halo stars which manifest as RGB stars similar to those of Scl-MM-Dw2. The foreground stars and background galaxies are mainly avoided by our defined selection boxes. Generally, dwarfs in PISCeS are identified using both techniques, although depending on the image quality one or the other method is superior on a case-by-case basis.  Details of our search methodology will be presented in a future work.

Scl-MM-Dw2 is very diffuse and is at the edge of two Megacam pointings. Thus, it is difficult to detect via visual inspection, but it is a clear dwarf candidate in our RGB maps (Figure \ref{map}).  Scl-MM-Dw2 is highly elongated, and points toward NGC~253 to the Southwest, with some signs of tidal extension.  Figure~\ref{CMD} shows our color-magnitude diagram (CMD) of Scl-MM-Dw2 within one half-light radius (see Section \ref{stellarpops}).  Due to uneven image quality, we cannot determine if Scl-MM-Dw2's extended structure directly connects with NGC~253's outer halo, but future {\it Hubble Space Telescope} (HST) Cycle 23 observations may shed light on this (PID: 14259; PI Crnojevi\'c).

\section{Properties of SCL-MM-DW2}\label{galaxy}

\subsection{Stellar Populations and Distance}\label{stellarpops}

\begin{figure*}
\centering
\includegraphics[angle=0,width=18cm]{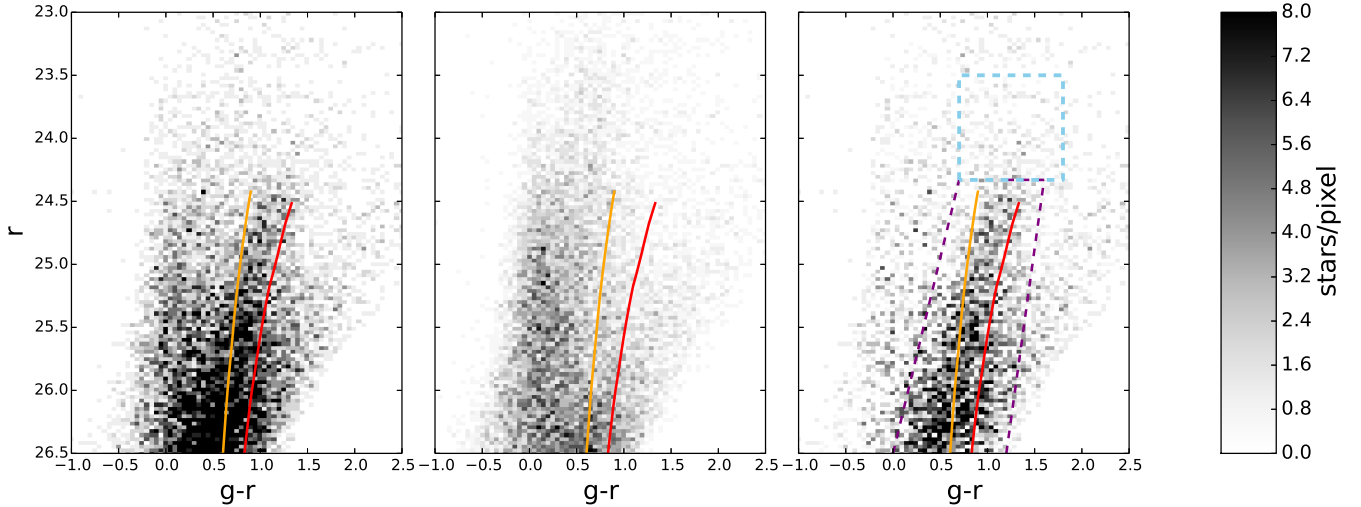}
\caption{De-reddened color-magnitude Hess diagrams (density CMD). Left panel: CMD showing the stars within the \Reff\ of the newly discovered dwarf galaxy. Middle panel: CMD showing the average stars within two background field regions, one in each of the Megacam fields where the galaxy lands, with the same area as in the left panel to account for possible small scale structure and radial variations in NGC~253's halo. Right panel: CMD showing the difference between the left and middle panels. Statistically, this CMD shows only stars within the newly discovered galaxy. The orange and red lines indicate the Padova isochrones for 12~Gyr and [Fe/H]~$=-2.3$~dex and $-1.1$~dex, respectively.
The purple and blue boxes indicate the selection boxes for RGB and AGB candidate stars, respectively. }\label{CMD}
\end{figure*}

We study the stellar content of the newly discovered Scl-MM-Dw2 by superposing old, metal-poor isochrones \citep[][hereafter Padova isochrones]{Marigo08,Girardi10} over the CMD of the galaxy. Figure \ref{CMD} shows the density CMD (a.k.a. Hess diagram) of an elliptical area within one half light radius (see \S~\ref{structure} for our structural parameter analysis). 

We estimate the distance to Scl-MM-Dw2 by using the tip of the red giant branch (TRGB) method \citep{Lee93,Salaris02,Rizzi07} which is based on the identification of the brightest metal-poor RGB stars present in the galaxy, as marked by a sharp change in the shape of the luminosity function. We apply a Sobel edge filter to the $r$ band luminosity function to look for this transition and restrict the color range to $0.8<g-r<1.2$ to minimize the contamination from foreground stars, background galaxies, and more metal rich stars from NGC~253's halo.

We find the TRGB at $r=24.46\pm0.18$. Using $M_r^{TRGB}=-3.01\pm0.10$ \citep[as determined for the SDSS $r$ band in][]{Sand14}
we obtain a distance modulus for Scl-MM-Dw2 of $m-M=27.47\pm0.21$. This distance is consistent with that of NGC~253 at the $\sim$1-$\sigma$ level \citep[assuming a distance modulus of 27.70$\pm$0.07 for NGC~253;][]{RadburnSmith11}. 

\begin{figure*}
\centering
\includegraphics[angle=0,width=8.5cm]{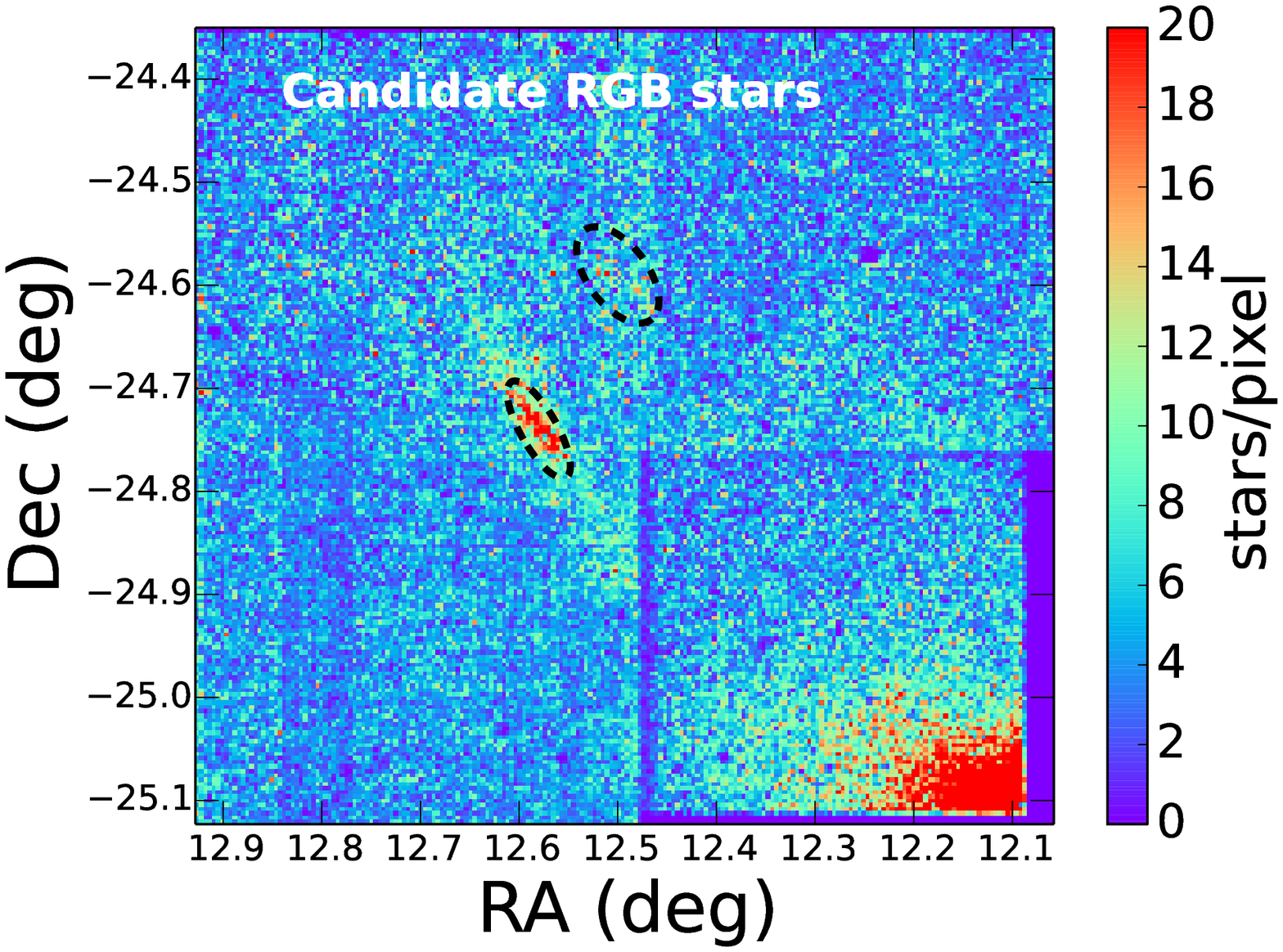}
\includegraphics[angle=0,width=8.5cm]{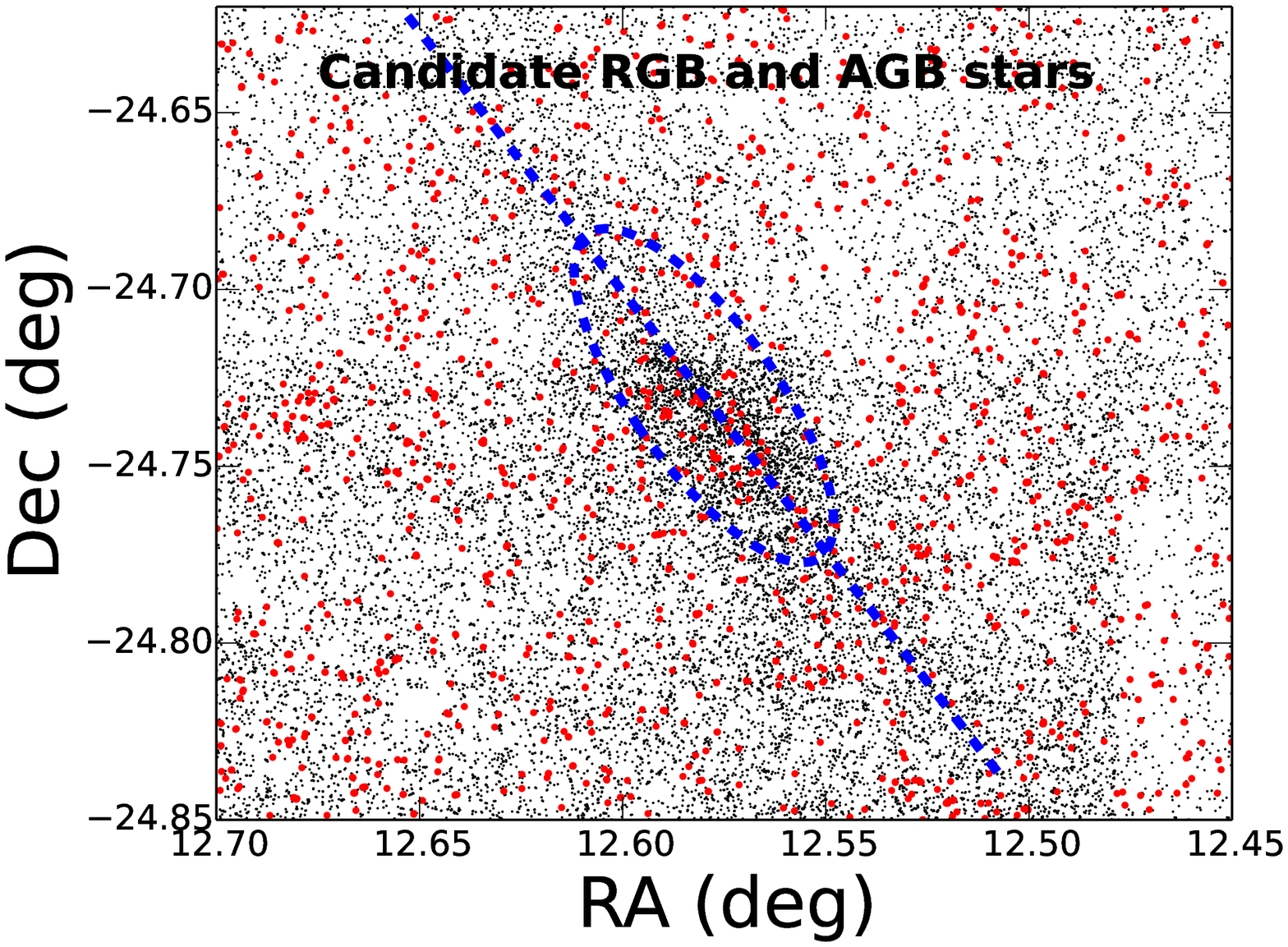}
\caption{Candidate RGB and AGB stars maps. These candidate stars are selected using the purple and blue boxes shown in Figure \ref{CMD}. Left panel: the black dashed ellipse in the overdensity region of RGB stars indicates the half-light radius and ellipticity of Scl-MM-Dw2.The black dashed ellipse to the Northwest of the galaxy indicates the position of a possible small overdensity of RGB stars. Right panel: zoom-in of Scl-MM-Dw2, with the spatial distribution of candidate RGB stars in black, and AGB stars in red. The ellipse is the same as in the left panel. The blue dashed line extends the major axis of the red ellipse to 2.5~\Reff. While Scl-MM-Dw2 is very prominent in the candidate RGB stars it appears to have only a few candidate AGB stars centrally concentrated. This suggests that the bulk of the stellar population is metal poor and old with a younger component in the center.}\label{map}
\end{figure*}

We use this distance modulus to place the theoretical Padova isochrones at the distance of Scl-MM-Dw2. Figure \ref{CMD} shows two isochrones that straddle the RGB ridgeline stars seen in the CMD, which correspond to an age of 12~Gyr and metallicities of [Fe/H]~$=-2.3$~dex and $-1.1$~dex. 
Figure \ref{map} shows the density map of candidate RGB stars and the stellar distribution of RGB and asymptotic giant branch (AGB) candidates. These maps are constructed by selecting all the stars within the boxes indicated in Figure \ref{CMD}.

 We estimate the number of AGB stars over the number of bright RGB stars (0.5~mag below the TRGB) and compare to models of different ages and metallicities. We find that this ratio decreases with the galaxy's radius ($0.20\pm0.04$ within the half-light radius, $r_h$, and $0.15\pm0.02$ within $2r_h$). This is consistent with a central young ($2-3$~Gyr) and metal poor ($-2.3<$~[Fe/H]~$<-1.1$) stellar population superposed on the overall old ($\sim12$~Gyr) and metal poor population. We also estimate the total stellar mass of the galaxy assuming a stellar mass-to-light ratio of 1.6 \citep[][see Table \ref{properties}]{Kirby13}. The stellar mass obtained is consistent with the value estimated by comparing the number of bright RGB stars with a modelled population of $10^6$~M$_{\odot}$.

\subsection{HI gas limits}

We investigated the possibility of HI gas associated with Scl-MM-Dw2.   Indeed, the HI Parkes All Sky Survey spectrum \citep[HIPASS;][]{Barnes01} exhibits a $\sim 3.7\sigma$ HI emission peak along the line of sight to Scl-MM-Dw2 with a heliocentric radial velocity of $\sim 275$~\kms\, suggesting a possible association with the dwarf. Seeking to confirm this tentative detection, we therefore obtained much deeper position-switched HI observations using Director's Discretionary time (program AGBT$\_$15B$\_$349) on the Green Bank Telescope (GBT; PI: Spekkens). The resulting spectrum has an rms noise $\sigma_{GBT} = 0.75$~mJy at a spectral resolution of 10~\kms, and is over an order of magnitude more sensitive than the corresponding HIPASS spectrum. 

We do not find any HI emission at $-700$~\kms~$\leq V_{hel} \leq -100$~\kms\ and $100$~\kms~$\leq V_{hel} \leq 1500$~\kms, whereas the range $-100$~\kms~$\leq V_{hel} \leq 100$~\kms\ is contaminated by Milky Way HI disk emission. The faint HIPASS feature is therefore a statistical fluke or an artifact of the bright HI emission from the nearby NGC~253 \citep{Lucero15}. Our non-detection implies that unless the radial velocity of Scl-MM-Dw2 puts it in the contaminated region of our GBT spectrum, any HI counterpart to Scl-MM-Dw2 has a 5$\sigma$, 15~\kms\ HI mass upper limit of $M_{HI}^{lim} = 1.2 \times 10^5$~M$_{\odot}$, and thus a gas fraction of $M_{HI}/L_V < 0.02$~\MLsun. Scl-MM-Dw2 is therefore gas-poor, similar to other dwarf spheroidal galaxies in the Local Volume \citep[e.g.][]{Grcevich09,Spekkens14}.

\subsection{Structure}\label{structure}

Figure \ref{map} shows the spatial distribution of candidate RGB and AGB stars. 
To the Northwest of Scl-MM-Dw2, with coordinates close to RA~$\sim12.5^{\circ}$ and Dec~$\sim -24.6^{\circ}$ there is a second ellipse away from the dwarf's main body that highlights a possible faint overdensity of stars. This possible overdensity coincides with the overlap region between different Megacam fields, and is a noisier area. Deeper photometry is necessary to confirm if these stars are true RGB stars and if they are related to Scl-MM-Dw2.
The candidate RGB stellar map clearly shows that Scl-MM-Dw2 is a very elongated galaxy that points towards the halo of NGC~253, which appears in the Southwest of the map. 

We measure the structural parameters of Scl-MM-Dw2 using the  maximum likelihood technique described by \citet{Martin08} and following the  implementation of \citet{Sand12}. We select all candidate RGB stars (utilizing the purple selection box in Figure \ref{CMD}) within a spatial box of $\sim$$15' \times 15'$ approximately centered on Scl-MM-Dw2. We fit the candidate RGB stars with  an exponential profile plus a constant background with the following free parameters: the central coordinates of the galaxy (RA$_0$,DEC$_0$),  position angle (PA), ellipticity ($\epsilon$), half-light radius ($r_h$), and the background surface density ($\Sigma_b$). The uncertainties in these parameters are estimated using 1000 bootstrap resamples of the data, and we report the $68\%$ confidence intervals. The measured structural parameters are presented in Table \ref{properties}.  

The surface brightness within the half light radius $\langle\mu_{h,V}\rangle$ is calculated from the total luminosity (see below, \S~\ref{lum}), \Reff, and ellipticity. The central surface brightness, $\mu_{0,V}$, is calculated using the conversion from $\langle\mu_{h,V}\rangle$ for an exponential profile from \citet{GrahamDriver05}. Scl-MM-Dw2 is large, elongated, and very diffuse, with $r_h\sim$3~kpc, $\epsilon=0.66$, and $\langle\mu_{h,V}\rangle=28.8$~mag~arcsec$^{-2}$.
The measured ellipticity is similar to the disrupting Sagittarius dwarf around the Milky Way \citep[MW;][]{McConnachie12}.  Along with the orientation and proximity of Scl-MM-Dw2 to NGC~253, this suggests that it too is being tidally disrupted, which we will discuss further in \S~\ref{discussion}.

\begin{table}
\begin{center}
\caption{Properties of Scl-MM-Dw2 \label{properties}}
{\renewcommand{\arraystretch}{1.}
\resizebox{8cm}{!} {
\begin{tabular}{l|c}
\hline \hline
Parameter         &  Value  \\
\hline
RA$_0$~(hh:mm:ss)   & 00:50:17.06$\pm 4.00''$       \\
DEC$_0$~(dd:mm:ss) & -24:44:58.58$\pm 7.30''$    \\
$m-M$~(mag)         & $27.47\pm0.21$ \\
$D$~(Mpc)               & $3.12\pm0.30$  \\
$M_V$~(mag)           & $-12.1\pm0.5$       \\
$r_h$~(arcmin)         & $3.24\pm 0.51$ \\
$r_h$~(kpc)                & $2.94\pm0.46$ \\
$\epsilon$               & $0.66\pm0.06$ \\
PA~(N to E; deg)       & $31.0\pm3.3$ \\
$\mu_{0,V}$~(mag~arcsec$^{-2}$) &  $26.5\pm0.7$   \\
$\langle\mu_{h,V}\rangle$~(mag~arcsec$^{-2}$) &   $28.8\pm0.6$  \\
M$_*$~(M$_{\odot}$) &  $1.0\pm0.4 \times 10^7$\\
$M_{HI}$~(M$_{\odot}$)  & $<1.2 \times 10^5$ \\
$M_{HI}/L_V$~(\MLsun) & $< 0.02$ \\
\hline
\end{tabular}
}}
\end{center}
\tablecomments{RA$_0$ and DEC$_0$ indicate the central coordinates of the galaxy in J2000. $M_V$ is measured in the AB system. $\mu_{0,V}$  and $\langle\mu_{h,V}\rangle$ are the central and effective surface brightness. These parameters are described in Section \ref{structure}.}
\end{table}

\subsection{Luminosity}\label{lum}

We estimate the total luminosity of Scl-MM-Dw2 via direct aperture photometry using an elliptical aperture with semimajor axis equal to the \Reff\ of the galaxy, semi-minor axis equal to $r_h\times(1-\epsilon)$, and the appropriate position angle. We estimate the flux within this aperture and subtract the estimated background by placing the same elliptical aperture randomly at different positions in the image. The resulting flux corresponds to the flux within the \Reff\ of Scl-MM-Dw2, which we then multiply by a factor of two to account for the total flux of the galaxy. 
The uncertainty was calculated based on the direct Poisson uncertainty in the Scl-MM-Dw2 aperture, as well as the scatter in measurements from the background apertures.

The final apparent magnitudes are $r=15.0\pm0.5$ and $g=15.9\pm0.5$; after applying our measured distance modulus and the filter transformations of \citet{Jester05} for stars with $R-I<1.15$, we derive $M_V$=$-12.1\pm0.5$ (see Table~\ref{properties}).

\subsection{Scaling Relations}

\begin{figure*}
\centering
\includegraphics[angle=0,width=8cm]{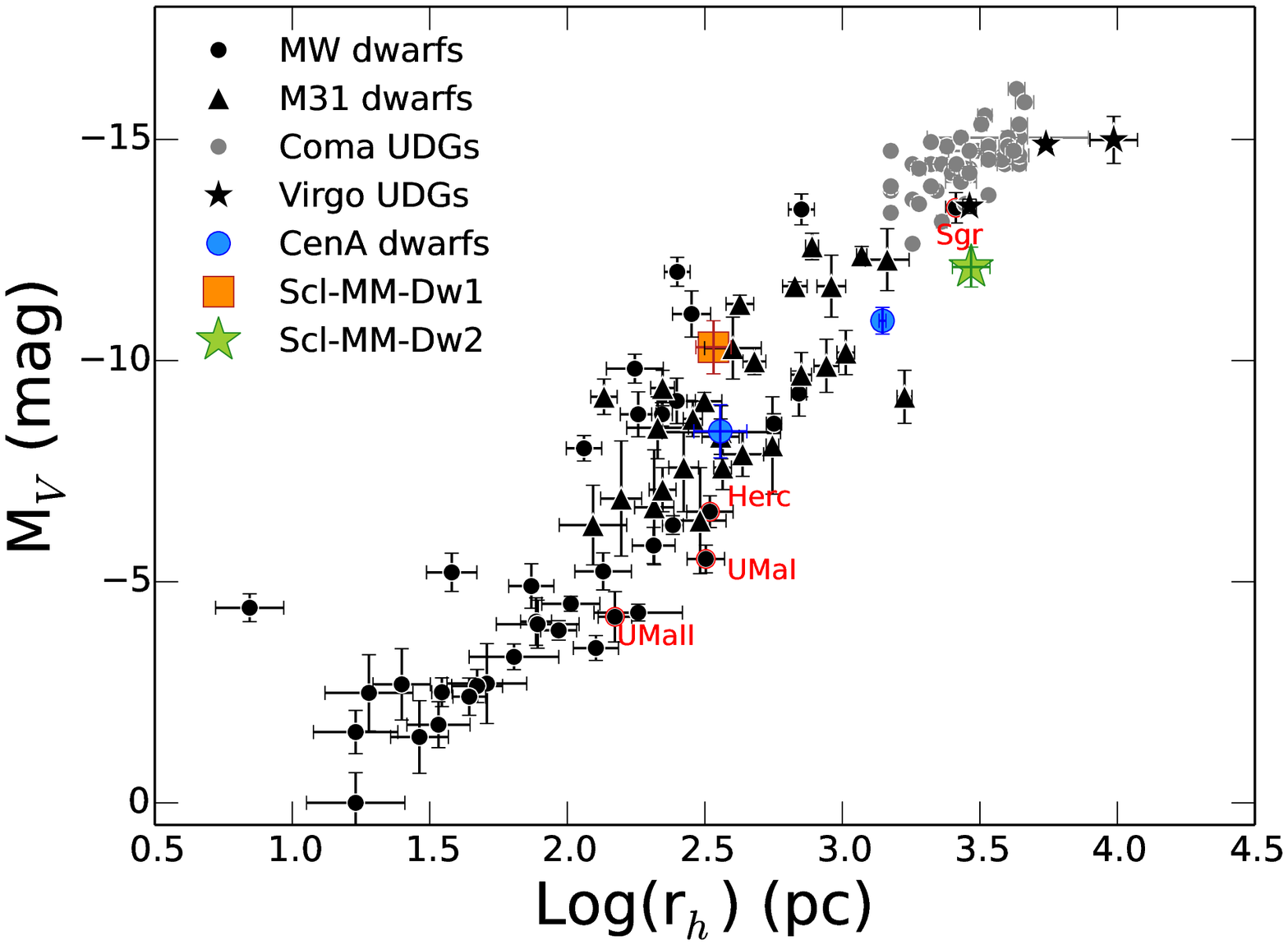}
\includegraphics[angle=0,width=8cm]{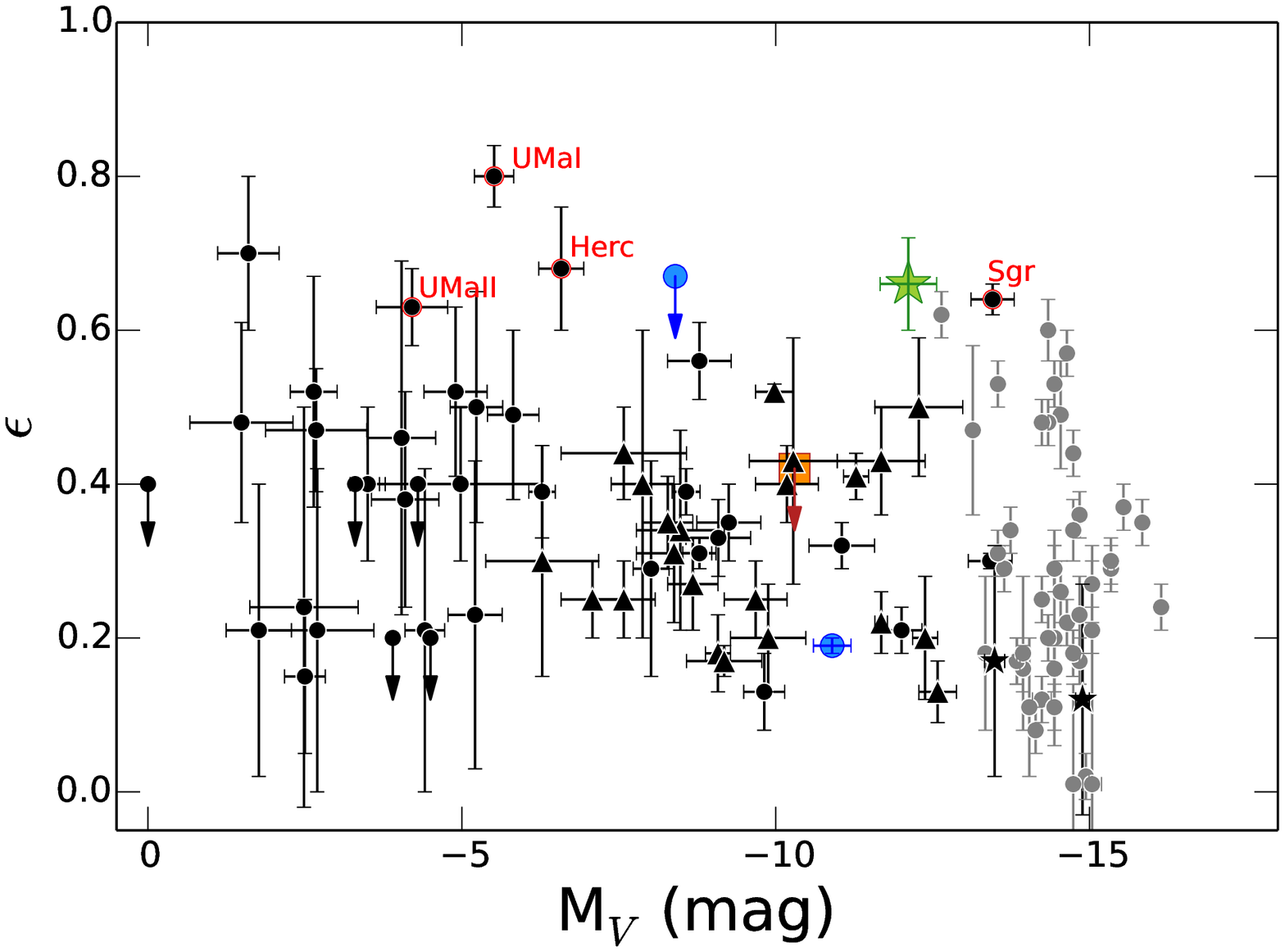}
\includegraphics[angle=0,width=8cm]{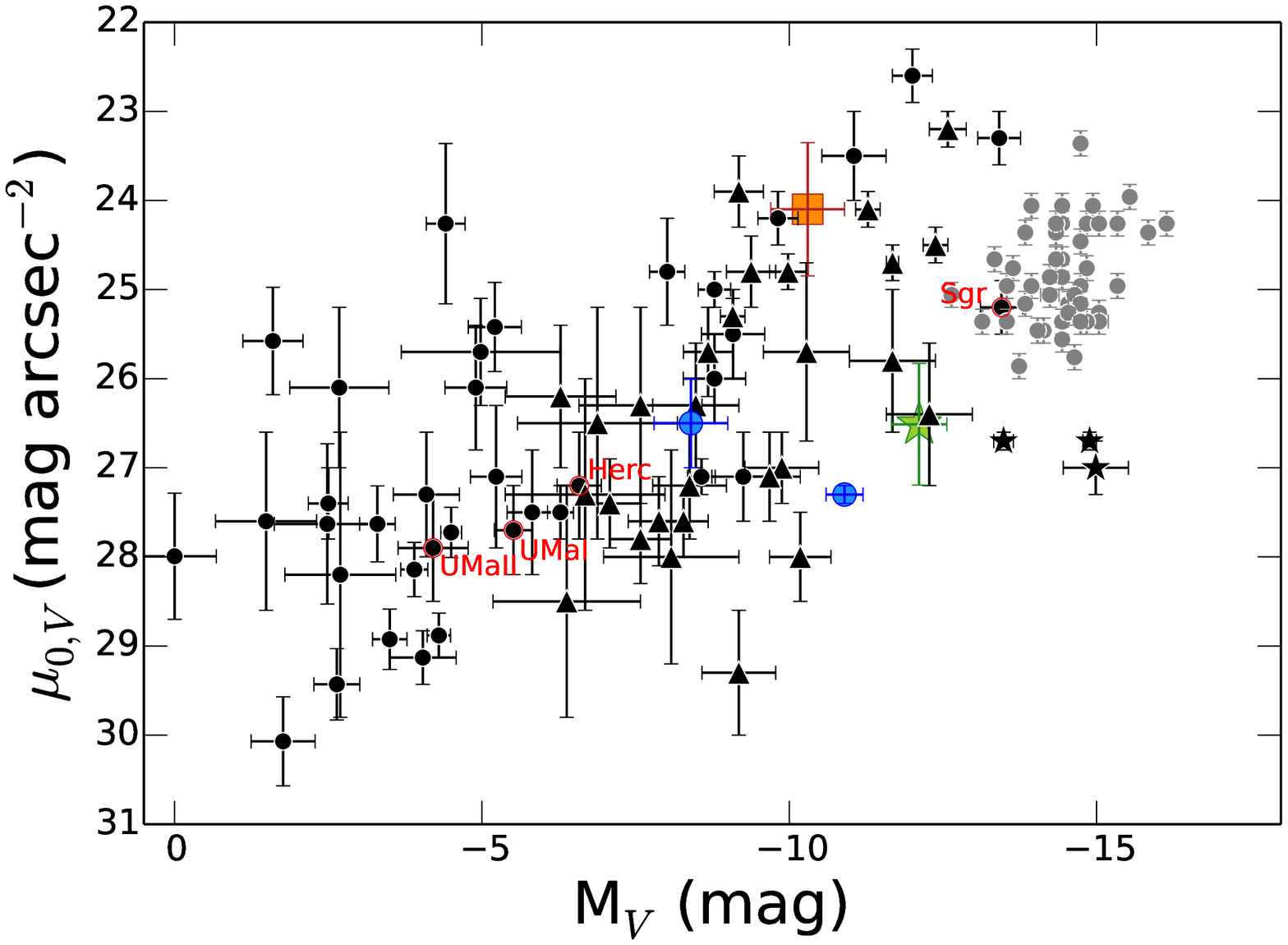}
\caption{Scaling relations for the Milky Way and M31 dwarf systems, along with recent discoveries around Centaurus~A, NGC~253, and the ultra diffuse galaxies (UDGs) in the Virgo and Coma clusters. From top to bottom and left to right are the size-luminosity relation, the ellipticity-luminosity relation, and the central surface brightness-luminosity relation. The black dots and triangles are the MW and M31 dwarf galaxies from \citet{McConnachie12,Martin13a,Martin13b,DES2015,Kim15a,Kim15b,Laevens15a,Laevens15b,Martin15}. Sagittarius, Hercules, Ursa~Major~I, and Ursa~Major~II are highlighted in red because they show evidence of tidal features. The black stars and grey dots are the UCDs found in the Virgo and Coma clusters, respectively \citep{Mihos15,vandokkum15}. The blue and orange squares are our discovered dwarf galaxies in the halo of Centaurus~A and NGC~253, respectively, as part of the PISCeS \citep{Crnojevic14,Sand14}. The green star is the newly discovered dwarf galaxy in the halo of NGC~253 presented in this work. This new galaxy, Scl-MM-Dw2, is among the largest and brightest dwarf satellites, while it has the lowest surface brightness and highest elongation for its luminosity. Scl-MM-Dw2 is likely being disrupted by NGC~253.}\label{scalingrelations}
\end{figure*}

Figure \ref{scalingrelations} shows the size-luminosity, ellipticity-luminosity, and surface brightness-luminosity scaling relations for Scl-MM-Dw2 in comparison with the dSphs in the Local Group \citep{McConnachie12,Martin13a,Martin13b,DES2015,Kim15a,Kim15b,Laevens15a,Laevens15b,Martin15}, our other 3 recent discoveries in the PISCeS survey \citep{Sand14,Crnojevic14}, and the ultra diffuse galaxies found in the Virgo and Coma clusters \citep{Mihos15,vandokkum15}. 

In the size-luminosity relation, Scl-MM-Dw2 is a slight outlier, being larger than most dwarf galaxies for its luminosity.  Meanwhile, it is amongst the most elongated faint dwarfs, with $\epsilon$=0.66$\pm$0.06, which is comparable to the disrupting Sagittarius dwarf and the ultrafaint galaxies Hercules, Ursa~Major~I and Ursa~Major~II \citep[e.g.][]{Okamoto08,Sand09,Munoz10}, all in the MW.  

Scl-MM-Dw2 stands out for its very low surface brightness, which is $\sim2$~mag~arcsec$^{-2}$ lower than Local Group dwarfs at a similar luminosity. 
Its very low surface brightness is only comparable to the slightly brighter ultra diffuse galaxies recently found in the Virgo cluster \citep{Mihos15}. 

\section{Discussion and Conclusions}\label{discussion}

We report the discovery of Scl-MM-Dw2, a dwarf galaxy with a luminosity of M$_V=-12.1$ located only $\sim 49'$ ($\sim 50$~kpc) in projection from NGC~253's center \citep[$\sim 3.6\times r_h^{\rm NGC253}$;][]{RC3}. We use the TRGB method to find a distance modulus of $m-M=27.47\pm0.21$ which places this galaxy at $3.12\pm0.30$~Mpc, consistent with that of NGC~253 within the uncertainty \citep{RadburnSmith11}. Thus, it is likely that Scl-MM-Dw2 is a satellite of NGC~253. 

We superpose Padova isochrones on the color-magnitude diagram centered in the galaxy and find that its stellar distribution can be explained with an old, $\sim 12$~Gyr, and metal poor, $-2.3<$~[Fe/H]~$<-1.1$, stellar population. In addition, we detect some AGB stars in the center of the galaxy that are consistent with a population of $2-3$~Gyr and $-2.3<$~[Fe/H]~$-1.1$. Thus, this dwarf has a central metal poor intermediate-age population superposed on the overall old stellar population. We do not detect any HI gas in emission in our GBT spectrum which constrains  the gas mass to an upper limit of $M_{HI}^{lim}=1.2\times 10^5$~M$_{\odot}$ and therefore a gas fraction $M_{HI}/L_V<0.02$~\MLsun. These properties place this new galaxy in the category of gas poor dwarf spheroidal galaxy.

We estimate the structural parameters of Scl-MM-Dw2 and compare them with those of other dSphs in the Local Group as well as from our own PISCeS program, and with the ultra diffuse galaxies found in the Virgo and Coma clusters. We find that Scl-MM-Dw2 is an outlier in all scaling relations. Its size is slightly larger than expected and it is extremely elongated and diffuse for its luminosity. The only other outlier on these relations is the Sagittarius dwarf galaxy in the Milky Way halo. However, while Sagittarius has a central star cluster \citep[M54;][]{Sarajedini95}, Scl-MM-Dw2 does not show any nucleus.
Moreover, the major axis of Scl-MM-Dw2 points towards the center of NGC~253 suggesting that it may have undergone a very close pericenter passage. 
In addition, we detect an overdensity of RGB stars on the Northwest and Southeast of the galaxy, which could be a tidal tail of Scl-MM-Dw2.

\citet{Klimentowski09} simulated a dwarf galaxy under the gravitational influence of a Milky Way analog and find that two kinds of tidal tails form: (1) the densest tidal tails, formed in the vicinity of the dwarf, are oriented towards the center of the host galaxy; and (2) the more diffuse tidal tails, formed on a much larger spatial scale and less likely to be detected due to their extremely low luminosity, are oriented along the orbit of the dwarf galaxy. 
\citet{Sand12} investigate the dependence of the angle between all Milky Way dSphs and the Galactic Center and find no clear correlation between their orientation and their elongation. Similar results have been found for M31 dSphs \citep{Salomon15}. 

The angle between the major axis of Scl-MM-Dw2 and the center of NGC~253 is only $\sim18^{\circ}$. This measurement is affected by projection effects, thus, the orientation of the galaxy alone is not enough evidence of this process. However, all these properties summarized above seem to point towards Scl-MM-Dw2 being disrupted by NGC~253.

\acknowledgments

E.T., D.J.S, J.D.S, and P.G. acknowledge the NSF grant AST-1412504. E.T. and P.G. are also supported by NSF grant AST-1010039 and D.J.S. by AST-1517649. K.S. acknowledges support from the Natural Sciences and Engineering Research Council of Canada (NSERC).

\bibliographystyle{aa}
\bibliography{references}{}

%%%%%%%%%%%%%%%%%%%%%%%%%%%%%%%%%%%%

\end{document}